\documentclass{article}
\usepackage[preprint]{spconf}
\usepackage{amsmath,graphicx}
\usepackage{hyperref}
\usepackage{cite,url,booktabs,subfig,bm,amssymb,multirow,xspace} 

\newcommand{\fo}{$F_0$\xspace}
\newcommand{\lfo}{$\log F_0$\xspace}

\newcommand{\bc}{\mathbf{c}}
\newcommand{\be}{\mathbf{e}}
\newcommand{\bx}{\mathbf{x}}

\newcommand{\bepsilon}{\bm{\epsilon}}

\newcommand{\periodnetV}{\textbf{PeriodNet~(voc)}\xspace}
\newcommand{\priorgradV}{\textbf{PriorGrad~(voc)}\xspace}
\newcommand{\periodgradV}{\textbf{PeriodGrad~(voc)}\xspace}

\newcommand{\periodnetM}{\textbf{PeriodNet~(ms+F0)}\xspace}
\newcommand{\priorgradM}{\textbf{PriorGrad~(ms+F0)}\xspace}
\newcommand{\periodgradM}{\textbf{PeriodGrad~(ms+F0)}\xspace}

\title{PeriodGrad: Towards Pitch-Controllable Neural Vocoder \\ Based on a Diffusion Probabilistic Model}
\name{Yukiya Hono, Kei Hashimoto, Yoshihiko Nankaku, and Keiichi Tokuda
\thanks{This work was supported by JSPS KAKENHI Grant Number JP22H03614, FOUNDATION OF PUBLIC INTEREST OF TATEMATSU, and CASIO SCIENCE PROMOTION FOUNDATION.}}
\address{Nagoya Institute of Technology, Nagoya, Japan}
\begin{document}
\ninept
\maketitle
\begin{abstract}
  This paper presents a neural vocoder based on a denoising diffusion probabilistic model (DDPM) incorporating explicit periodic signals as auxiliary conditioning signals. Recently, DDPM-based neural vocoders have gained prominence as non-autoregressive models that can generate high-quality waveforms. The neural vocoders based on DDPM have the advantage of training with a simple time-domain loss. In practical applications, such as singing voice synthesis, there is a demand for neural vocoders to generate high-fidelity speech waveforms with flexible pitch control. However, conventional DDPM-based neural vocoders struggle to generate speech waveforms under such conditions. Our proposed model aims to accurately capture the periodic structure of speech waveforms by incorporating explicit periodic signals. Experimental results show that our model improves sound quality and provides better pitch control than conventional DDPM-based neural vocoders.
\end{abstract}
\copyrightnotice{
  \begin{tabular}{l}
    \copyright~2024 IEEE.
    Personal use of this material is permitted.
    Permission from IEEE must be obtained for all other uses, in any current or \\future media, including reprinting/republishing this material for advertising or promotional purposes, creating new collective works, \\for resale or redistribution to servers or lists, or reuse of any copyrighted component of this work in other works.
  \end{tabular}
}

\vspace{-1mm}
\section{Introduction}
\vspace{-2mm}
\label{sec:intro}

A neural vocoder is a deep neural network (DNN) that generates speech waveforms from acoustic features and have been used in various speech applications, including speech synthesis~\cite{shen-2018-natural}, singing voice synthesis~\cite{hono-2021-sinsy}, and voice conversion.
The success of these applications depends heavily on the capabilities of the neural vocoder, such as generated sound quality, inference speed, and controllability.

There are several types of neural vocoders, such as autoregressive (AR)~\cite{oord-2016-wavenet,mehri-2017-samplernn,kalchbrenner-2018-efficient,valin-2019-lpcnet} and non-AR ones~\cite{oord-2018-parallel,prenger-2019-waveglow,ping-2020-waveflow,yamamoto-2020-parallel,kumar-2019-melgan,kong-2020-hifi}.
Notably, non-AR neural vocoders leveraging generative adversarial networks (GANs)~\cite{goodfellow-2014-generative} have gained popularity in generating high-quality speech waveforms at high speed~\cite{yamamoto-2020-parallel,kumar-2019-melgan,kong-2020-hifi}.
These methods are usually challenging to train with only adversarial loss and need to be combined with multiple auxiliary losses with weighting parameters, leading to complicated training procedures. 

In recent advances in image generation, denoising diffusion probabilistic models (DDPMs)~\cite{song-2019-generative,ho-2020-denoising,song-2020-score,kingma-2021-variational} have emerged as promising generative models that outperforms traditional GAN-based models~\cite{dhariwal-2021-diffusion,rombach-2022-high}.
Several studies~\cite{chen-2020-wavegrad,kong-2020-diffwave} have successfully incorporated DDPMs into neural vocoders, which can be trained with a simple time-domain loss function while achieving generated sound quality comparable to AR neural vocoders.
However, DDPMs involve an iterative denoising process during inference, resulting in a trade-off between performance and speed.
Later studies have proposed data-dependent adaptive priors~\cite{lee-2022-priorgrad,koizumi-2022-specgrad}, improved modeling frameworks~\cite{okamoto-2021-noise,takahashi-2023-hierarchical}, and better training strategies~\cite{chen-2022-infergrad} to reduce the number of iterations while maintaining sound quality.

Since neural vocoders are data-driven approaches, they present challenges in controllability compared with conventional signal-processing-based vocoders~\cite{morise-2016-world}.
In particular, the controllability of the fundamental frequency (\fo) is an essential issue for neural vocoders in practical applications such as speech and singing voice synthesis.
As an extension to GAN-based neural vocoders, several methods inputting sinusoidal signals corresponding to the pitch of the speech waveform as explicit periodic signals have been proposed to achieve superior pitch controllability~\cite{hono-2021-periodnet,yoneyama-2023-source,matsubara-2023-harmonic}.
Another effect of using periodic signals is the capability to generate speech waveforms with higher sampling rates, such as 48 kHz, without increasing the model size or changing the model structure~\cite{hono-2021-periodnet}.
Such pitch-controllable, high-sampling-rate speech waveform generation models are in demand for professional use cases such as music production.
Nevertheless, despite this, DDPM-based neural vocoders suitable for these practical use cases have not been sufficiently investigated.
Tackling these challenges will broaden the range of applications of DDPM-based neural vocoders.

In this paper, we introduce a novel DDPM-based neural vocoder conditioned by explicit periodic signals, following previous pitch-robust neural vocoders.
The proposed model is based on PriorGrad~\cite{lee-2022-priorgrad}, which can generate speech waveforms with reasonable inference cost.
The experimental results show that the proposed model improves the sound quality of speech waveforms at high sampling rates and \fo controllability.

\vspace{-1mm}
\section{DDPM-based neural vocoder}
\vspace{-2mm}
\label{sec:ddpm}

Let $\bx_0=(x_1, x_2, \ldots, x_N)$ be a speech waveform corresponding to the acoustic feature sequence $\bc=(\bc_1, \bc_2, \ldots, \bc_K)$, where $N$ is the number of samples of the speech waveform and $K$ is the number of frames of the acoustic feature.
A neural vocoder is defined as a DNN that generates a sample sequence of the speech waveform $\bx_0$ corresponding to the acoustic feature sequence $\bc$.

\vspace{-1mm}
\subsection{Overview of DDPM}
\vspace{-1mm}

A DDPM is a deep generative model defined by two Markov chains: the \emph{forward} and \emph{reverse} processes.
The \emph{forward process} gradually diffuses the data $\bx_0$ to standard noise $\bx_T$ as follows:
\begin{align}
  q(\bx_{1:T}|\bx_0) =& \prod_{t=1}^{T} q(\bx_{t}|\bx_{t-1}), \label{eq:forward-process}
\end{align}
where $T$ is the number of steps of DDPMs, and $q(\bx_{t}|\bx_{t-1}) = \mathcal{N}(\bx_t;\sqrt{1-\beta_t}\bx_{t-1},\beta_t\mathbf{I})$ is a transition probability that adds small Gaussian noise in accordance with a predefined noise schedule $\{\beta_1,...,\beta_T\}$.
This formulation enables us to sample $\bx_t \sim q(\bx_t|\bx_0)$ at an arbitrary timestep $t$ in a closed form as
\begin{align}
  \bx_t &= \sqrt{\bar{\alpha}_t} \bx_0 + \sqrt{1-\bar{\alpha}_t} \bepsilon,
\end{align}
where $\alpha_t = 1- \beta_t$, $\bar{\alpha}_t=\prod_{s=1}^t \alpha_s$, and $\bepsilon \sim \mathcal{N}(\mathbf{0},\mathbf{I})$.

The \emph{reverse process} is a denoising process that gradually generates data $\bx_0$ from standard noise $p(\bx_T)$ as follows:
\begin{align}
  p_{\theta}(\bx_{0:T}) &= p(\bx_T) \prod_{t=1}^T p_{\theta}(\bx_{t-1}|\bx_t), \label{eq:reverse-process}
\end{align}
where $p_{\theta}(\bx_{t-1}|\bx_t)$ is modeled by a DNN with parameters $\theta$.
As both forward and reverse processes have the same function form when $\beta_t$ is small, the transition probability of the reverse process is parameterized as $p_{\theta}(\bx_{t-1}|\bx_t) \!=\! \mathcal{N}\left(\bx_{t-1}; \bm{\mu}_\theta(\bx_t,t),\gamma_t\mathbf{I}\right)$, where $\gamma_t \!=\! \frac{1 - \bar{\alpha}_{t-1}} {1 - \bar{\alpha}_t} \beta_t$ and $\gamma_1 \!=\! 0$.
The mean $\bm{\mu}_\theta(\bx_t,t)$ is defined as
\begin{align}
  \bm{\mu}_\theta (\bx_t,t) &= \frac{1}{\sqrt{a_t}} \left(\bx_t - \frac{\beta_t}{\sqrt{1 - \bar{\alpha}_t}} \bepsilon_\theta(\bx_t, \bc, t)\right),
\end{align}
where $\bepsilon_\theta(\bx_t, \bc, t)$ is a DNN for predicting noise contained in $\bx_t$.

A DDPM can be regarded as a latent variable model with $\bx_{1:T}$ as the latent variable.
The model $\bepsilon_\theta(\bx_t, \bc, t)$ can be optimized by maximizing the evidence lower bound (ELBO) of the log-likelihood $p(\bx_0)$.
However, DDPM-based neural vocoders~\cite{chen-2020-wavegrad,kong-2020-diffwave} generally use a simplified loss $L_{\mathrm{DDPM}}(\theta)$, following \cite{ho-2020-denoising}, as
\begin{align}
  L_{\mathrm{DDPM}}(\theta) &= \mathbb{E}_q \left[ || \bepsilon - \bepsilon_\theta(\bx_t, \bc, t) ||^2_2 \right], \label{eq:loss-simple}
\end{align}
where $||\cdot||_p$ is the $L_p$ norm.

\vspace{-1mm}
\subsection{PriorGrad}
\vspace{-1mm}

The pioneer DDPM-based neural vocoders, WaveGrad~\cite{chen-2020-wavegrad} and DiffWave~\cite{kong-2020-diffwave}, require over 200 iterations to achieve sufficient quality comparable to AR neural vocoders.
PriorGrad introduces an adaptive prior $\mathcal{N}(\bm{0}, \bm{\Sigma}_\bc)$, where the diagonal variance $\bm{\Sigma}_\bc$ is computed from $\bc$ as $\bm{\Sigma} = \mathrm{diag}[(\sigma^2_1, \sigma^2_2, \ldots, \sigma^2_N)]$, where $\sigma^2_n$ represents the power at the $n$-th sample obtained by interpolating the normalized frame-level energy calculated from $\bc$.
The loss function is also modified to use the Mahalanobis distance in accordance with $\bm{\Sigma}_\bc$, as
\begin{align}
  L_{\mathrm{Prior}}(\theta) &= \mathbb{E}_q \left[ || \bepsilon - \bepsilon_\theta(\bx_t, \bc, t) ||^2_{\bm{\Sigma}_\bc^{-1}} \right], \label{eq:loss-priograd}
\end{align}
where $||\bx||^2_{\bm{\Sigma}^{-1}} = \bx^\top \bm{\Sigma}^{-1} \bx$.
Intuitively, as the power envelope of the adaptive prior is closer to that of the target speech waveform than that of the standard Gaussian prior, PriorGrad achieves faster model convergence and inference with better denoising performance.

\vspace{-1mm}
\section{Proposed method: PeriodGrad}
\vspace{-2mm}

Speech waveforms are strongly autocorrelated signals, a characteristic that makes them inherently different from other tasks where DDPMs have been successful, such as image generation.
Existing DDPM-based neural vocoders need to learn the periodic structure of speech in an entirely data-driven manner, which may limit the flexibility of \fo control during inference.
Additionally, it may also be challenging to generate periodic speech even with a limited amount of training data and high sampling rates.
Using explicit periodic information may be helpful for DDPM-based neural vocoder in generating speech waveforms.

We propose PeriodGrad, a DDPM-based neural vocoder that leverages explicit periodic signals as conditions.
In PeriodGrad, the extended noise estimation model $\bepsilon_\theta(\bx_t, \bc, \be, t)$ denoises the noise from the input signal $\bx_t$ conditioned on the auxiliary feature $\bc$ and the periodic signal $\be=[\be_1,\be_2,\ldots,\be_N]$.
PeriodGrad uses the sine-based periodic signal, which consists of sample-level signals concatenated with sine waves and voiced/unvoiced (V/UV) signals as the periodic signal $\be$, as in the previous study~\cite{hono-2021-periodnet}.

Any model structure can be used, by simply introducing an additional condition embedding layer.
PeriodGrad can be trained using the same training criterion as conventional DDPM-based neural vocoders, such as Eq.~\eqref{eq:loss-simple} or Eq.~\eqref{eq:loss-priograd}.
According to PriorGrad~\cite{lee-2022-priorgrad}, we adopt the energy-based adaptive prior, and the model is trained using the following loss function:
\begin{align}
  L_{\mathrm{Period}}(\theta) &= \mathbb{E}_q \left[ || \bepsilon - \bepsilon_\theta(\bx_t, \bc, \be, t) ||^2_{\bm{\Sigma}_\bc^{-1}} \right]. \label{eq:loss-priograd-proposed}
\end{align}

\vspace{-2mm}
\section{Experiment}
\vspace{-2mm}
\label{sec:exp}

\subsection{Experimental conditions}
\vspace{-1mm}

\begin{figure*}[t]
  \centering
  \subfloat{\includegraphics[width=0.95\hsize]{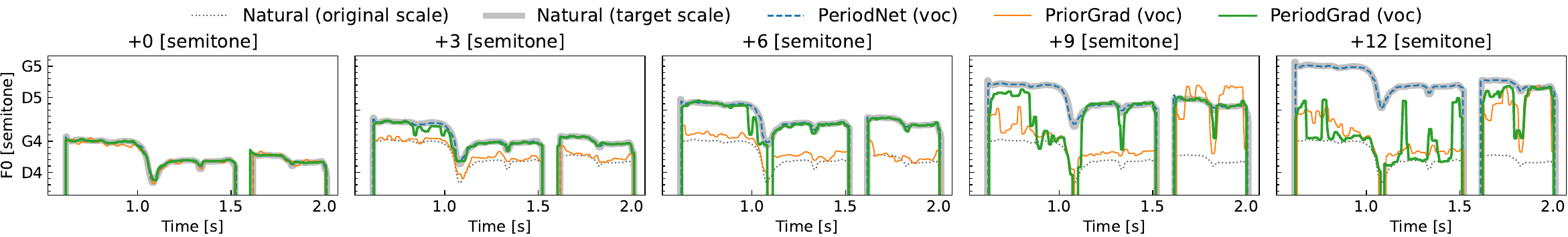}}
  \\
  \vspace{-1mm}
  \subfloat{\includegraphics[width=0.95\hsize]{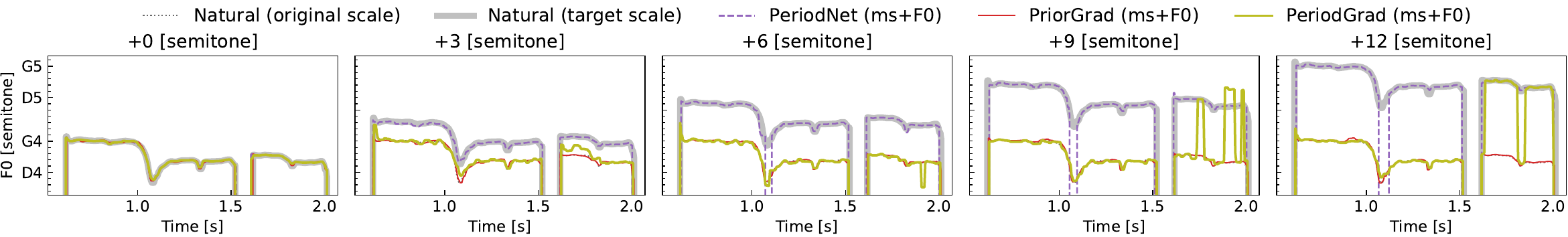}}
  \vspace{-1mm}
  \caption{\fo contours of the natural and generated singing voices.}
  \label{fig:lf0-all}
  \vspace{-3.5mm}
\end{figure*}

\begin{figure}[t]
  \centering
  \vspace{-2mm}
  \subfloat[\fo-RMSE \label{fig:obj-lf0}]{\includegraphics[height=3.45cm]{{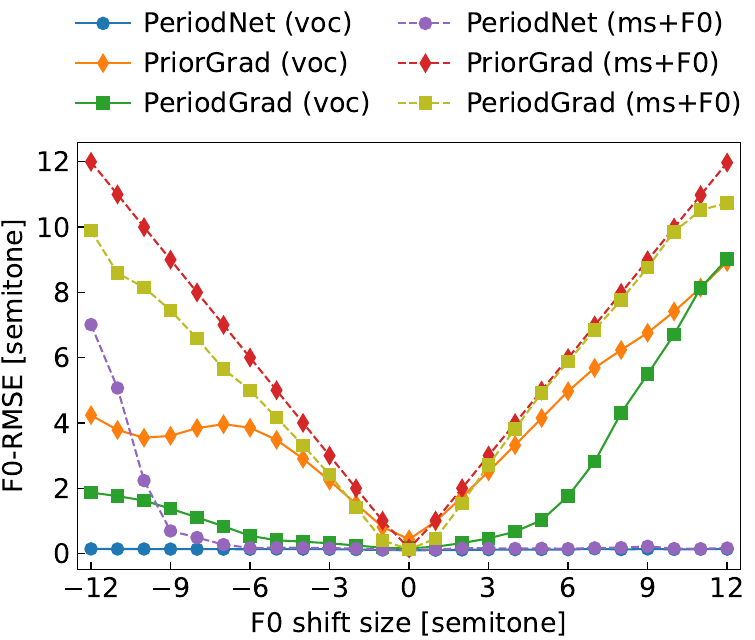}}}
  \hspace{1mm}
  \subfloat[V/UV-ER \label{fig:obj-vuv}]{\includegraphics[height=3.45cm]{{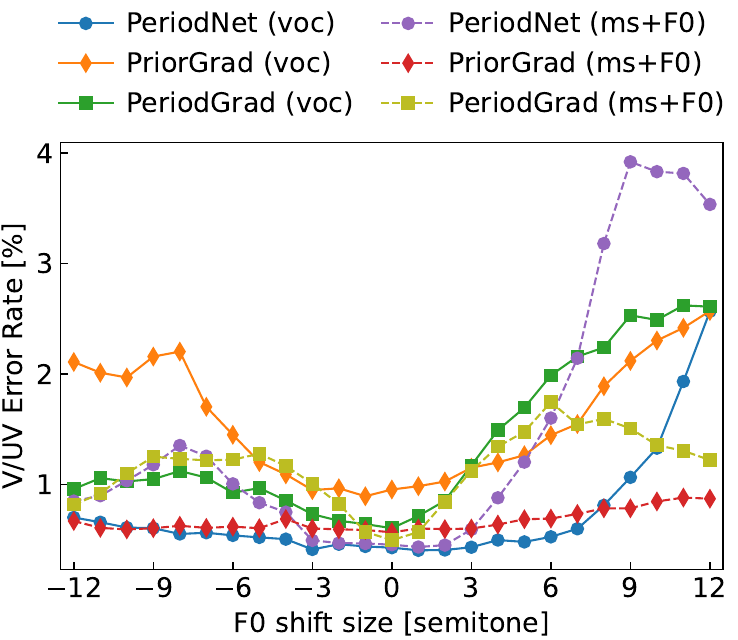}}}
  \hspace{2mm}
  \vspace{-1mm}
  \caption{Results of objective evaluation for \fo accuracy.}
  \label{fig:obj-all}
  \vspace{-3.5mm}
\end{figure}

We conducted experiments using 70 Japanese children's songs by one single female singer.
Sixty songs (approx. 70 min.) were used for training, and the remaining ten songs (approx. 6 min.) were used for testing.
The sampling frequency of the audio waveform was 48kHz, and the quantization bit was 16 bits.
We used two types of acoustic feature sets: \textbf{voc}: 50-dimensional mel-cepstral coefficients, a continuous \lfo value, 25-dimensional aperiodicity, and a V/UV binary flag, and \textbf{ms+F0}: 80-dimensional log mel-spectrograms, a continuous \lfo value, and a V/UV binary flag.
Note that \textbf{ms+F0} is the same configuration employed in several singing voice synthesis approaches~\cite{chen-2020-hifisinger}.
Mel-cepstral coefficients were extracted by WORLD~\cite{morise-2016-world}.
Mel-spectrograms were extracted with 2048-point fast Fourier transform using a 25-ms Hanning window.
Voting results from three different \lfo extractors were used to reduce the impact of extraction errors~\cite{sawada-2016-nitech}.
The \lfo was interpolated before being fed into neural vocoders.
All feature vectors were extracted with a 5-ms shift, and the features were normalized to have zero mean and unit variance before training.
The explicit periodic signal used as the input to the neural vocoders was generated based on the glottal closure instants extracted from the natural waveform during training and based on non-interpolated \lfo during inference.

We compared PeriodGrad with two neural vocoders: PriorGrad, as a DDPM-based baseline model~\cite{lee-2022-priorgrad}, and PeriodNet, as a pitch-controllable GAN-based model~\cite{hono-2021-periodnet}.
These methods were trained with two auxiliary feature sets: \textbf{voc} and \textbf{ms+F0}.

In PriorGrad, we used the same model architecture with 30 layers of non-causal dilated convolutions with three dilation cycles as in the original settings~\cite{lee-2022-priorgrad}, except that the upsampling scale was adjusted to a 5 ms frame shift.
The number of iterations during training and inference was set to 50 and 12, respectively.
The noise schedule was set to \texttt{linspace(1e-4, 0.05, 50)} during training and [0.0001, 0.0005, 0.0008, 0.001, 0.005, 0.008, 0.01, 0.05, 0.08, 0.1, 0.2, 0.5] during inference, by following the official implementation\footnote{\url{https://github.com/microsoft/NeuralSpeech/tree/master/PriorGrad-vocoder}}.
In the case of \textbf{ms+F0}, the normalized energy was calculated according to the original paper~\cite{lee-2022-priorgrad}.
In the case of \textbf{voc}, the normalized energy was derived from the impulse response calculated from the mel-cepstrum coefficients.

In PeriodGrad, we added a fully-connected layer into each block of non-causal dilated convolution in PriorGrad to embed the periodic signal and performed training and inference under the same conditions as in PriorGrad.

In PeriodNet, we used the PeriodNet parallel model denoted as PM1 in~\cite{hono-2021-periodnet}, which consists of a periodic and aperiodic generator.
A sine wave and V/UV signal were used as the periodic input signal of the periodic generator.
The model architecture and training configuration were the same as in~\cite{hono-2021-periodnet}.
The generator in PeriodNet was trained using multi-resolution short-time Fourier transform (STFT) loss and adversarial loss.
The discriminator in PeriodNet adopted a multi-scale structure in the same configuration as~\cite{hono-2021-periodnet}.

\vspace{-1mm}
\subsection{Objective evaluation}
\vspace{-1mm}
\label{sec:obj-eval}

The root mean square error (RMSE) of \lfo (\fo-RMSE) [semitones] and the V/UV error rate (V/UV-ER) [\%] were used to evaluate the pitch accuracy of generated waveforms objectively.
We evaluated the normal copy-synthesis settings and the copy-synthesis with \lfo shifting by $-12$ to $+12$ semitones.

Figures~\ref{fig:lf0-all} and~\ref{fig:obj-all} show examples of \lfo extracted from the generated waveforms and the results of the objective evaluation.
From Fig.~\ref{fig:obj-lf0}, in most cases, it can be seen that \periodgradV has better accuracy in reproducing the given \lfo than \priorgradV.
This result indicates that using explicit periodic signals improves the \fo controllability in the DDPM-based neural vocoders, similar to GAN-based ones.
However, even with \periodgradV, the \fo-RMSE worsens significantly when the input \fo is shifted upward by six semitones or more.
In addition, the \fo-RMSE did not reach the level of \periodnetV for any shift amount.
Compared with PeriodNet, which deterministically generates periodic components from explicit periodic signals, PeriodGrad, which employs multiple sampling at inference under the DDPM framework, may have found it more challenging to generate waveforms with proper periodic structures corresponding to explicit periodic signals.

Incidentally, both \priorgradM and \periodgradM could not reproduce the target \lfo when the shifted \lfo was fed into these methods, as shown in Fig~\ref{fig:lf0-all}, resulting in a significant \fo-RMSE deterioration.
\periodnetM also has a distinctly worse \fo-RMSE than \periodnetV when the \lfo was downward-shifted by more than ten semitones.
The mel-spectrogram contains the pitch information of speech, unlike the WORLD features.
Even if \fo had been explicitly used as an auxiliary feature, the neural vocoder would have modeled the speech waveform by focusing on the pitch information in the mel-spectrogram instead of given \fo.
We hypothesize there are two reasons why the explicitly given \fo tends to be ignored:
1)~Due to the difficulty of \fo extraction, there are extraction errors such as octave confusion and V/UV detection error in the extracted \fo.
2)~The unvoiced regions in the extracted \fo are linearly interpolated before being fed into the neural vocoder as a continuous feature.
In these cases, there is no direct relationship between \fo and the periodic structure of the waveform, which confuses the model.
In contrast, since these problems do not exist in the pitch information embedded in the mel-spectrogram, the models tend to trust the pitch information embedded in the mel-spectrogram instead of the explicitly given \fo.

It can also be seen from Fig.~\ref{fig:obj-vuv} that when the input \fo is largely shifted upward, the V/UV-ERR becomes worse. This is because when the neural vocoder generates a waveform whose pitch is outside the range of the training data, the generated waveform becomes noisy, and the voice tends to crack, making it difficult to perform proper \fo extraction in such cases.

\begin{figure*}[t]
  \centering
  \vspace{-2mm}
  \subfloat[Normal \label{fig:mos-normal}]{\includegraphics[height=2.88cm]{{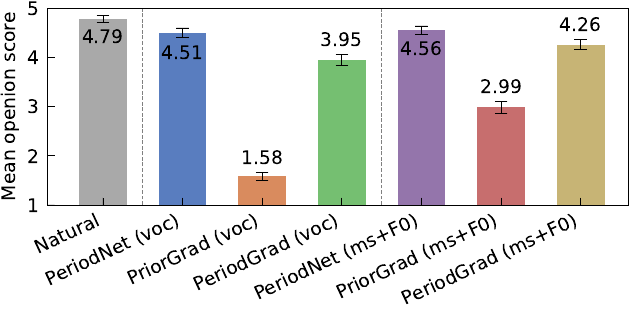}}}
  \hspace{1mm}
  \subfloat[High pitch ($+3$ semitone) \label{fig:mos-up}]{\includegraphics[height=2.88cm]{{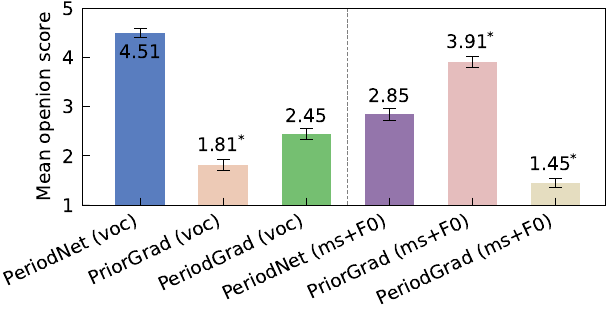}}}
  \hspace{1mm}
  \subfloat[Low pitch ($-3$ semitone) \label{fig:mos-down}]{\includegraphics[height=2.88cm]{{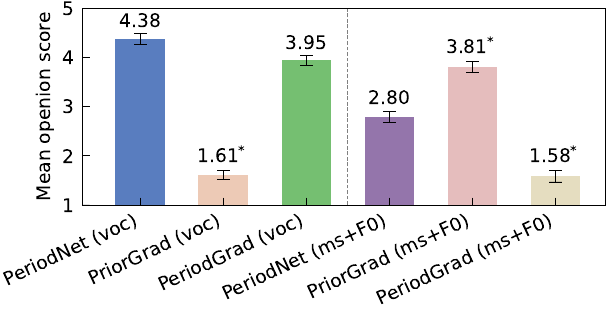}}}
  \vspace{-2mm}
  \caption{
    Results of subjective evaluation with 95\% confidence intervals.
    The methods annotated with * have insufficient pitch control performance.
    These methods are impractical, even if the subjective rating of sound quality could be better.
  }
  \label{fig:mos-all}
\end{figure*}

\begin{figure*}[t]
  \centering
  \includegraphics[width=0.98\hsize]{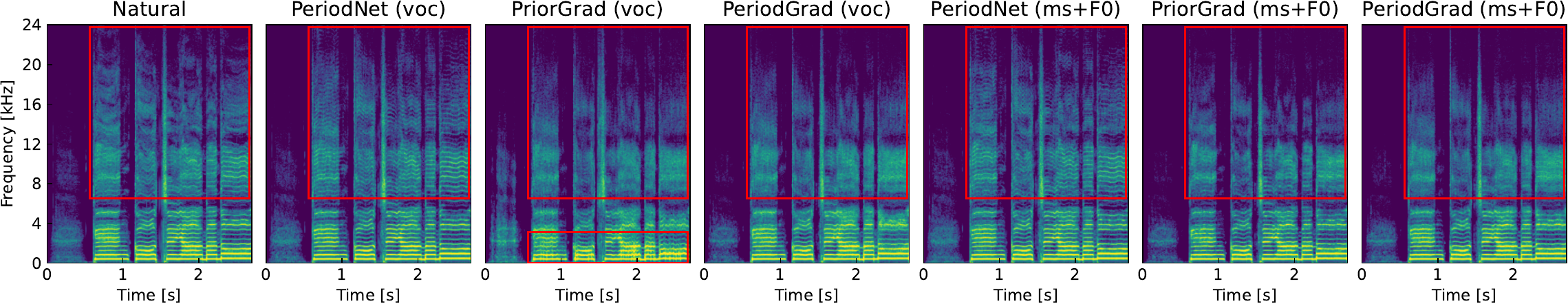}
  \vspace{-2mm}
  \caption{Spectrograms of the natural and generated singing voices.}
  \label{fig:spec-all}
  \vspace{-3.5mm}
\end{figure*}

\vspace{-1mm}
\subsection{Subjective evaluation}
\vspace{-1mm}
\label{sec:sub-eval}

We performed 5-scale mean opinion score (MOS) tests\footnote{Audio samples are available at the following URL: \url{https://www.sp.nitech.ac.jp/~hono/demos/icassp2024/}} to evaluate the quality of the generated singing voice waveforms.
In these experiments, samples were generated by each model conditioned on three different scales of \lfo: original \lfo, upward-shifted \lfo by 3 semitones (300 cents), and downward-shifted \lfo by 3 semitones (-300 cents).
Thirteen participants evaluated 10 phrases randomly selected from 10 songs in the test data and evaluated a total of six methods, combining feature sets \textbf{voc} and \textbf{ms+F0} for each of \textbf{PeriodNet}, \textbf{PriorGrad}, and \textbf{PeriodGrad}.
These listening tests were conducted separately.
In the experiment with the original \fo scale, the natural waveform \textbf{Natural} was also used for comparison.

The results of the subjective evaluation are presented in Fig.~\ref{fig:mos-all}.
Examples of spectrograms of the generated waveforms are also shown in Fig.~\ref{fig:spec-all}.
In the original \fo scale, the proposed \periodgradV significantly outperformed \priorgradV, as shown in Fig.~\ref{fig:mos-normal}.
The spectrogram of the waveform generated by \priorgradV showed an unnatural fluctuation at the low-frequency range, as shown in the highlighted boxes at the bottom of the \priorgradV in Fig.~\ref{fig:spec-all}, which decreased the quality of the generated speech waveform.
On the other hand, since the generated waveforms of \periodgradV did not show such degradation, its MOS value was significantly improved compared with \priorgradV, which suggests that the explicit periodic signal contributed significantly to the improvement of the quality of the generated speech waveform even in the DDPM-based neural vocoders.
However, \periodgradV is still not as good as PeriodNet in terms of the generated speech quality.
From Fig.~\ref{fig:spec-all}, it can be seen that PriorGrad and PeriodGrad have a low quality of generating harmonic components contained in the natural waveform above 6~kHz. 
This indicates that there is still room for improvement in the quality of the generated 48~kHz sampled waveform.

On the other hand, \periodnetM, \priorgradM, and \periodgradM showed better MOS scores than \periodnetV, \priorgradV, and \periodgradV, respectively, as shown in Fig.~\ref{fig:mos-normal}.
Notably, in \priorgradM, the unnatural fluctuation in the low-frequency range was not observed unlike the spectrogram of \priorgradV.
In addition, the quality of \periodgradM also approached that of \periodnetV and \periodnetM.
Using the mel-spectrogram rather than vocoder parameters extracted using the WORLD vocoder improved the sound quality of the generated waveform significantly.

We discuss the results of the case of \lfo shifting in Fig.~\ref{fig:mos-up} and Fig.~\ref{fig:mos-down}.
First, \priorgradV, \priorgradM, and \periodgradM were impractical since the pitch of generated sounds was not shifted properly, as mentioned in section~\ref{sec:obj-eval}.
In particular, \priorgradM showed a high score; however, this is not valuable.
Since \priorgradM ignores the given shifted \lfo inputs and consistently generates a waveform like the same as in the normal case, there is no conspicuous degradation due to pitch shifting.
Hence, its subjective score was easily higher than most other comparisons, showing degraded sound quality due to pitch shifting.
While \periodgradV showed better performance than \priorgradV, \periodgradV did not reach \periodnetV.
We found that the upward-shifted waveforms generated by \periodgradV sometimes contained noise.
Another noteworthy point is that the MOS score of \periodgradV decreased slightly with the \lfo downward shift and substantially with the \lfo upward shift, compared with \periodnetV.
\periodgradV, with multiple sampling in the DDPM inference process, may not be robust to \lfo shifting compared to the \periodnetV.
Incidentally, \periodgradM showed the lowest score for both the upward and downward \fo shifting cases.
In \periodgradM, when \fo is shifted, both the components corresponding to the shifted and original \fo appear in the generated waveform.
This phenomenon indicates that \periodgradM also utilizes pitch information embedded in the mel-spectrogram, which does not change with \fo shifting, along with the \fo and the periodic signal.
Note that a similar phenomenon sometimes occurred when high-pitch waveforms were generated in \periodgradV.
This result suggests that the mel-cepstrum also retains some information correlated with \fo.
Appropriate disentanglement of pitch and spectrum parameters is a promising direction for future work.

\vspace{-2mm}
\section{Conclusion}
\vspace{-2mm}

We proposed a DDPM-based neural vocoder called \emph{PeriodGrad} that uses an explicit periodic signal as an additional condition.
The proposed model can generate a speech waveform while considering the periodic structure of the speech waveform explicitly in the reverse process of the DDPM.
The experimental results showed that PeriodGrad achieved better sound quality and \fo controllability than the conventional DDPM-based neural vocoder in the task of generating 48-kHz singing voice waveforms.
While there are still challenges in certain scenarios, PeriodGrad would mark a significant step towards providing the ability to control the pitch of the output waveform in DDPM-based neural vocoders.

Future work includes conducting experiments using various kinds of waveforms, such as speech and music, to investigate the performance of the proposed model.
In addition, disentangling pitch information from spectral parameters is an important issue in building a pitch-controllable DDPM-based neural vocoder with better performance and robustness.

\bibliographystyle{IEEEtran}
\bibliography{references}

\end{document}